# Ideal Weyl fermions and double Kagome bands in a series of distorted armchair-type all-$sp^2$ carbon networks

Yun-Yun Bai[1], Yan Gao[1]*, Weikang Wu[2], Yong Liu[1]*, and Zhong-Yi Lu[3,4]

[1]State Key Laboratory of Metastable Materials Science and Technology & Hebei Key Laboratory of Microstructural Material Physics, School of Science, Yanshan University, Qinhuangdao 066004, China

[2]Key Laboratory for Liquid-Solid Structural Evolution and Processing of Materials (Ministry of Education), Shandong University, Jinan, Shandong 250061, China

[3]School of Physics and Beijing Key Laboratory of Opto-electronic Functional Materials & Micro-nano Devices, Renmin University of China, Beijing 100872, China

[4]Key Laboratory of Quantum State Construction and Manipulation (Ministry of Education), Renmin University of China, Beijing 100872, China

*Corresponding authors: yangao9419@ysu.edu.cn; yongliu@ysu.edu.cn

# Abstract

The study of the Weyl fermions and Kagome bands has recently attracted significant attention in condensed matter physics. However, realizing of perfect Weyl semimetals and double Kagome bands remains challenging. Here, we report a new class of distorted armchair-type fully $sp^2$-hybridized carbon networks, termed DACN-$n$. The DACN-$n$ family is characterized by only six pairs of Weyl points on the $k_z = 0$ plane near the Fermi level. We calculated the chirality of these Weyl points and found that each carries a topological charge of $\pm 1$. Notably, DACN-5 exhibits a double Kagome band, where the Weyl points arise from the intersection between Dirac-type bands in the two sets of Kagome bands. The structural stability of DACN-$n$ is confirmed by phonon spectra, *ab initio* molecular dynamics simulations, and elastic constant calculations. Additionally, we investigated the surface states of the (001) surface and found that its nontrivial topological properties are attributed to the Fermi arcs connecting a pair of Weyl points with opposite chirality. We also simulated x-ray diffraction patterns to guide experimental synthesis. Our findings not only present a new family of three-dimensional carbon allotropes, but also provide a unique opportunity for exploring ideal Weyl fermions and double Kagome bands.

# 1. Introduction

Topological semimetals (TSMs) have been one of central topics in condensed matter physics and materials science for over a decade [1-3]. TSMs can be classified into three categories based on the dimensionality of their band crossing points within the Brillouin zone (BZ) [4]: zero-dimensional (0D) nodal point semimetals [5-9], one-dimensional (1D) nodal line semimetals [10-12], and two-dimensional (2D) nodal surface semimetals [13-15]. Among these, nodal point semimetals are further divided into subtypes, including twofold degenerate Weyl semimetals (WSMs) [5-7], fourfold degenerate Dirac semimetals (DSMs) [8, 9], and three-, six- and eightfold degenerate semimetals [16, 17]. Particularly, WSMs have attracted significant attention due to their exotic and rich physical properties [18, 19]. In WSMs, the low-energy quasiparticles around the Weyl point can be described by the Weyl equation. The formation of these Weyl points requires the breaking of either time-reversal (*T*) symmetry and/or inversion (*P*) symmetry [7, 20]. Weyl points carry a non-zero Chern number (topological charge) and occur in pairs with opposite chirality (±), acting as sources (+) and sinks (−) of the Berry curvature field in momentum space. Most WSMs reported to date are found in materials containing heavy elements with strong spin-orbit coupling (SOC), such as $Y_2Ir_2O_7$ [5], $HgCr_2Se_4$ [20], $TaAs$ [7], TaIrTe$_4$ [21], and MoTe$_2$ [22], etc. While strong SOC can lead to nontrivial band topologies and enable the realization of Weyl fermions, it can also induce energy gaps at the band crossings, potentially affecting the stability of the Weyl points [1].

However, in materials with weak SOC effects, such as carbon, which has a negligible SOC effect, the spin can be treated as a dummy degree of freedom [10, 23]. Once there are bands intersect to form crossing points near the Fermi level of the spatial inversion-broken three-dimensional (3D) carbon allotropes, it is very likely to form the "spinless" Weyl fermion without considering the effect of SOC on it. Moreover, carbon is one of the most attractive elements in the periodic table, able to form compounds with other elements as well as a rich variety of carbon allotropes from 0D to 3D, and the usually structurally stable carbon phases are mainly derived from $sp$, $sp^2$, $sp^3$ and their hybrid bonds [24, 25], such as 0D fullerenes [26], 1D carbon nanotubes [27], 2D graphyne/graphene [28, 29] and 3D graphite/diamond [30], etc. Within them, the discovery of graphene not only sparked a wave of exploring 2D materials, but also laid a solid foundation for the exploration of 3D carbon materials [31, 32]. Nevertheless, WSMs with ideal WPs resulting only from two non-degenerate bands crossing near the Fermi level are still relatively rare [33-35]. Therefore, 3D carbon allotropes with broken space-inversion symmetry can serve as alternatives ideal material platform to explore WSMs that are obviously different from those with sizable SOC effects [36, 37].

On the other hand, the study of Kagome lattice and Kagome bands has received increasing attention [38-40]. The Kagome lattice, characterized by the 2D network of corner-sharing triangles, presents both spin and orbital frustration among neighboring atoms, which can facilitate the generation of flat bands and give rise to various novel many-body states [41], In such systems, Dirac-type bands merge with a nontrivial flat band to form the Kagome bands. Nevertheless, a single Kagome band at the Fermi level does not simultaneously exhibit the physical properties of the Dirac fermions in the Dirac band and the strongly correlated fermions in the flat band. Remarkably, the emergence of the double Kagome bands near the Fermi level addresses this issue [42]. The double Kagome bands exhibit a unique electron transport phenomenon, and its electron mobility is more pronounced than that of materials with a single Kagome band. To the best of our knowledge, there have been no reports in the literature concerning the double Kagome bands in 3D materials.

In this work, we predict a series of distorted armchair-type (DA) fully $sp^2$-hybridized carbon networks (CN), named as DACN-*n*. The DACN-*n* family is composed of flat armchair graphene nanoribbons (FAGNRs) with different width (*n*) and the triangulated distorted armchair graphene nanoribbons (DAGNRs). The structural stability of DACN-*n* is confirmed by phonon spectrum, *ab initio* molecular dynamics simulations, and elastic constant calculations. By investigating the electronic properties of the DACN-*n* family, we find that it comprises two non-degenerate bands that intersect to form twofold degenerate points near the Fermi level, resulting in six perfect pairs of Weyl points with Chen number $C = \pm 1$ distributed on the $k_z = 0$ plane. It is worth noting that DACN-5 has double Kagome bands, and its WPs are derived from the intersection between Dirac-type bands in the two sets of Kagome bands. The surface states of the (001) surface confirm that its nontrivial topological properties come from the Fermi arcs connecting a pair of Weyl points with opposite chirality. Our work not only provides a new class of members to the family of 3D carbon allotropes, it also offers a desirable platform for the study of perfect WeyI fermions and double Kagome bands.

## 2. Computational methods

The electronic structures of DACN-*n* (*n* = 1-5) were investigated by using the projective augmented plane wave method [43] as implemented in the VASP code [44] in the framework of density functional theory (DFT). The exchange correlation functional was treated using the generalized gradient approximation (GGA) of the Perdew-Burke-Ernzerhof (PBE) type [45]. The kinetic energy cutoff of the plane wave basis was set to 500 eV. The $6 \times 6 \times 10$, $4 \times 4 \times 10$, $4 \times 4 \times 10$, $4 \times 4 \times 10$, and $2 \times 2 \times 10$ *k*-point meshes [46] were taken for the Brillouin zone (BZ) sampling of the DACN-*n* (*n* = 1-5) structure, respectively. The energy and force convergence criteria were set to be $10^{-6}$eV and $10^{-3}$eV/Å, respectively. The phonon spectrum calculations of DACN-*n* (*n*=1-5) were carried out using the PHONOPY package [47] within finite displacement method by using $3 \times 3 \times 4$, $3 \times 3 \times 4$, $2 \times 2 \times 4$, $2 \times 2 \times 4$, and $2 \times 2 \times 3$ supercells, respectively. The thermal stability of DACN-*n* was evaluated with the *ab-initio* molecular dynamics (AIMD) simulations in a canonical ensemble with a Nose-Hoover thermostat [48]. The elastic stiffness tensor $C_{ij}$ were determined by using strain-stress method [49]. The topological properties of DACN-*n* family were analyzed with the Wannier90 [50] and WannierTools codes[51].

## 3. Results and Discussion

The crystal structures of DACN-*n* family are shown in Fig. 1 and Fig. S1 in the Supplementary Material (SM). The DACN-*n* family is composed of the triangulated

DAGNRs (blue atoms) and FAGNRs (grey atoms) linked together and differs in that these structures have FAGNRs with varying widths ($n$ = 1-5) [see Fig. 1(b, e) and Fig. S1(b, e, h) in the SM]. Considering that DACN-$n$ possesses similar crystal structures and electronic properties, we present only the structures with widths $n$ = 1 and $n$ = 5 [see Fig. 1(b) and 1(e)] for presentation in the main text. Detailed information on the remaining widths can be found in the SM. The DACN-$n$ family exhibits two distinct space groups: P6$_4$22 (No. 181) for odd $n$ and P6$_1$22 (No. 178) for even $n$ [see Table I]. Both groups share the same point group $D_6$. In the DACN-$n$ family, all carbon atoms are $sp^2$ hybridized. The optimized lattice parameters for DACN-1 are a = b = 7.75 Å and c = 4.59 Å, with a primitive cell containing 18 carbon atoms [see Fig. 1(c)] occupying two Wyckoff positions: 12k (0.444, 0.821, 0.854) and 6f (0.500, 0.500, 0.005). The bond lengths range from 1.43 to 1.48 Å, which is shorter than the band length in diamond (1.54 Å). The bond length connecting DAGNRs and FAGNRs is 1.45 Å, comparable to that of graphene (1.42 Å). The bond angles range from 105.2° to 133.2°.

For DACN-5, the optimized lattice parameters are a = b = 15.89 Å and c = 4.38 Å, with a primitive cell consisting of 42 carbon atoms occupying four Wyckoff positions: 12k (0.415, 0.593, 0.004), 12k (0.385, 0.737, 0.852), 12k (0.456, 0.545, 0.500), and 6f (0.500, 0.500, 0.001) [see Fig. 1(f)]. The bond lengths vary from 1.40 Å to 1.46 Å, with the bond length of 1.44 Å connecting the DAGNRs and FAGNRs, which is in close agreement with that of graphene (1.42 Å). The minimum bond angle in DACN-5 is 106.9°, which is close to that of diamond (109.5°), while the maximum bond angle of 133.9° occurs at the junction of the DAGNRs and FAGNRs. The Wyckoff position of DACN-$n$ ($n$ = 1-5) summarized in Table S1 of the SM. The purple dashed region in Fig. 1(d) and 1(g) indicates that the DACN-$n$ family can be constructed of various widths $n$ as its basic building blocks. Condensing these building blocks into a point result in the formation of a Kagome lattice in the 2D plane [see Fig. 1(h)], enabling the appearance of Kagome bands.

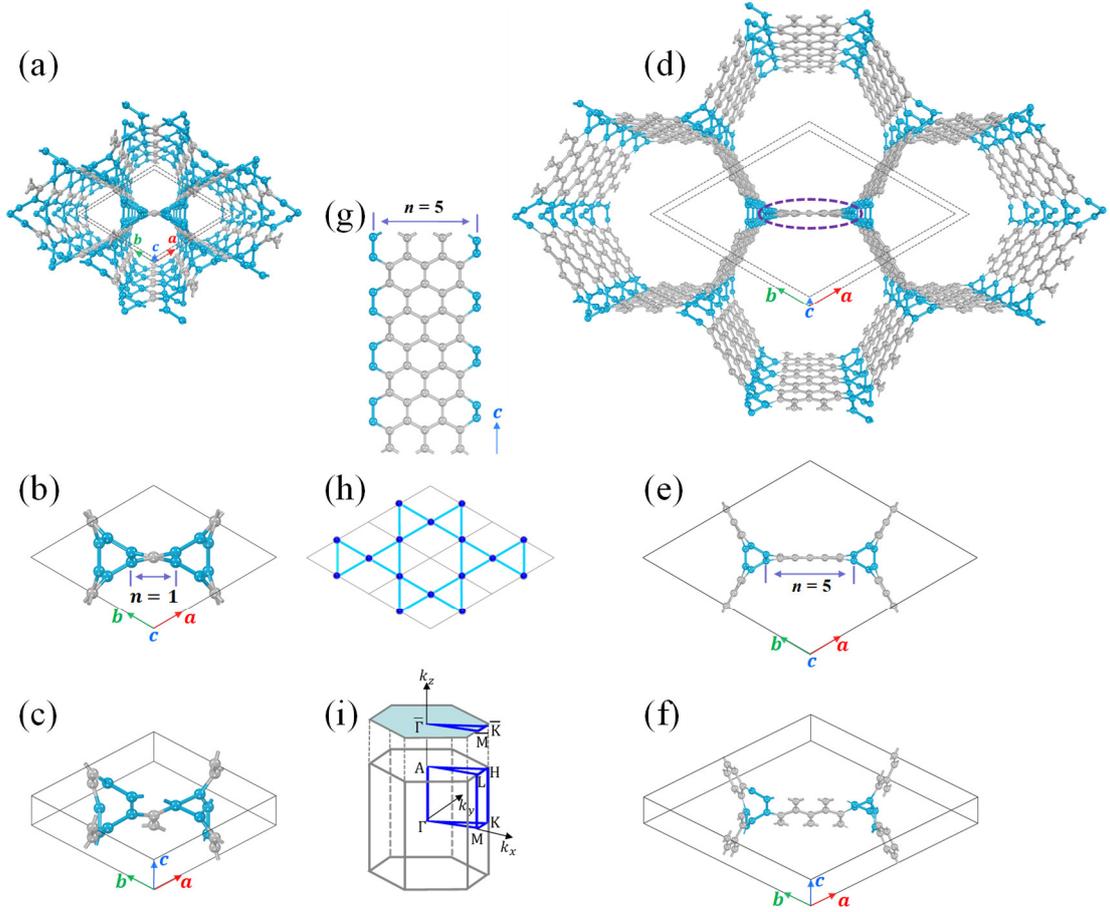

FIG. 1. Crystal structures and Brillouin zone (BZ) of DACN-1 and DACN-5. (a-c) The perspective, top and side views of DACN-1. (d-f) The perspective, top and side views of DACN-5, respectively. The basic structural unit of DACN-$n$ is marked by the purple dashed region in panel (d). (g) The side view of the purple dashed section after a 90° upward rotation. (h) The Kagome lattice in the 2D plane. (i) The bulk Brillouin zone (BZ) of DACN-$n$ and projected 2D BZ of the corresponding (001) surface.

To investigate the structural properties of DACN-$n$, we compared it with several typical carbon allotropes (see Table 1), such as Diamond, T-carbon [52], CHC-2 [53, 54], fco-$C_6$ [37], bco-$C_{16}$ [55], CNW-(3, 2) [56], CKL [57], and IGN [58]. We found that the total energy ($E_{tot}$) of DACN-$n$ decreases with the increasing $n$, with $E_{tot}$ for DACN-4 at -8.86 eV/C, comparable to -8.88 eV/C for the experimentally synthesized CHC-2. Meanwhile, the energy of DACN-$n$ approaches that of diamond (-9.09 eV/C), indicating favorable energetic stability. Additionally, DACN-$n$ exhibits a low bulk modulus; the bulk moduli of the DACN-n structures ($n \geq 3$) are $\leq$ 135.13 GPa, smaller than those of other carbon allotropes but larger than the bulk modulus of glass (35-55 GPa). This lower bulk modulus aligns with DACN-$n$'s low density, as its porous structure leads to reduced carbon density, which is dependent on pore size and inversely proportional to $n$. Notably, the density of DACN-2 (1.39 g/$cm^3$) is close to that of CHC-2 (1.29 g/$cm^3$), attributed to their structural similarities and porous characteristics, both of which make them suitable candidates for gas storage applications. Furthermore, DACN-$n$ ($n \geqslant 4$) possesses a density of $\leqslant 0.99$ g/$cm^3$, lower than that of water, thereby enhancing its potential for applications in aerospace and cosmic materials.

TABLE I. The space groups, lattice parameters, band lengths, densities, total energies ($E_{tot}$), bulk moduluses, and topological properties of DACN-$n$ ($n$ = 1-5), T-carbon, fco-$C_6$, bco-$C_{16}$, CHC-2, CKL, IGN, CNW-(3, 2), and diamond structures. WSM, NLSM, and TM refer to Weyl semimetal, nodal line semimetal, and topological metal, respectively.

| Structure | Space groups | Lattice parameters(Å) | | | Band lengths (Å) | Density (g/$cm^3$) | $E_{tot}$ (eV/C) | Bulk moduluses (GPa) | Properties |
|---|---|---|---|---|---|---|---|---|---|
| | | a | b | c | | | | | |
| DACN-1 | $P6_422$ | 7.75 | 7.75 | 4.59 | 1.43-1.48 | 1.72 | -8.60 | 199.7 | WSM |
| DACN-2 | $P6_122$ | 9.40 | 9.40 | 4.51 | 1.41-1.47 | 1.39 | -8.68 | 161.3 | WSM |
| DACN-3 | $P6_422$ | 11.61 | 11.61 | 4.43 | 1.40-1.48 | 1.16 | -8.80 | 135.5 | WSM |
| DACN-4 | $P6_122$ | 13.74 | 13.74 | 4.41 | 1.40-1.47 | 0.99 | -8.86 | 117.3 | WSM |
| DACN-5 | $P6_422$ | 15.89 | 15.89 | 4.38 | 1.40-1.46 | 0.87 | -8.91 | 103.1 | WSM |
| T-carbon | $I4_1/amd$ | 5.31 | 5.31 | 5.31 | 1.42,1.50 | 1.51 | -7.92 | 155.5 | Insulator |
| fco-$C_6$ | $F222$ | 8.84 | 6.82 | 3.35 | 1.41-1.49 | 2.37 | -8.46 | 270.4 | WSM |
| bco-$C_{16}$ | $Imma$ | 4.90 | 4.90 | 4.90 | 1.39-1.47 | 2.52 | -8.67 | 315.0 | NLSM |
| CHC-2 | $P6/mmm$ | 10.12 | 10.12 | 2.43 | 1.41, 1.45, 1.49 | 1.29 | -8.88 | 152.1 | TM |
| CKL | $P6_3/mmc$ | 4.46 | 4.46 | 2.53 | 1.50,1.53 | 2.75 | -8.81 | 321.4 | Insulator |
| IGN | $CMCM$ | 4.33 | 4.33 | 2.47 | 1.41,1.43 | 2.59 | -8.99 | 308.8 | NLSM |
| CNW-(3, 2) | $IMMA$ | 6.18 | 6.18 | 2.20 | 1.35-1.55 | 1.58 | -9.00 | 187.7 | NLSM |
| Diamond | $Fd\bar{3}m$ | 3.56 | 3.56 | 3.56 | 1.54 | 3.55 | -9.09 | 422.5 | Insulator |

To verify the stability of the DACN-$n$ family, we calculated the phonon spectra and found no soft phonon modes throughout the entire Brillouin zone (BZ) [see Figs. 2(a-b) and S2(a-c)], indicating that the DACN-$n$ family is dynamically stable. Additionally, AIMD simulations were performed, showing that the structure of DACN-$n$ remains intact after relaxation at 600 K for 8 ps, with energy fluctuations centering around the equilibrium position, further confirming its thermal stability [see Fig. 2(c-d) and Fig. S2(d-f)]. We also calculated the elastic constants for DACN-$n$, finding the five independent constants $C_{11}$, $C_{12}$, $C_{13}$, $C_{33}$, and $C_{44}$ for DACN-1 to be 229.55, 143.76, 104.16, 612.69, and 112.03 GPa, respectively. For DACN-5, the values were $C_{11} = 91.51$ GPa, $C_{12} = 92.38$ GPa, $C_{13} = 51.44$ GPa, $C_{33} = 343.18$ GPa, and $C_{44} = 54.91$ GPa. These values satisfy the mechanical stability criteria of the hexagonal crystal system [59]: $C_{11} > |C_{12}|$, $2C_{13}^2 < C_{33}(C_{11} + C_{12})$, and $C_{44} > 0$, thereby confirming the mechanical stability of DACN-$n$. To further investigate the energetics, we calculated the total energies versus volume curves for DACN-$n$ and eight typical carbon allotropes [see Fig. 3], including T-carbon, CHC-2, fco-$C_6$, bco-$C_{16}$, CKL, IGN, diamond, and graphite. It can be observed that the minimum total energy of DACN-$n$ decreases gradually with increasing $n$. The energy stability of DACN-4 is comparable to that of the experimentally synthesized CHC-2, which is consistent with previous analyses [see Table I]. Notably, DACN-$n$ is not only lower in energy than the theoretically predicted fco-$C_6$, but also significantly lower than the experimentally synthesized T-carbon, suggesting its potential for experimental synthesis.

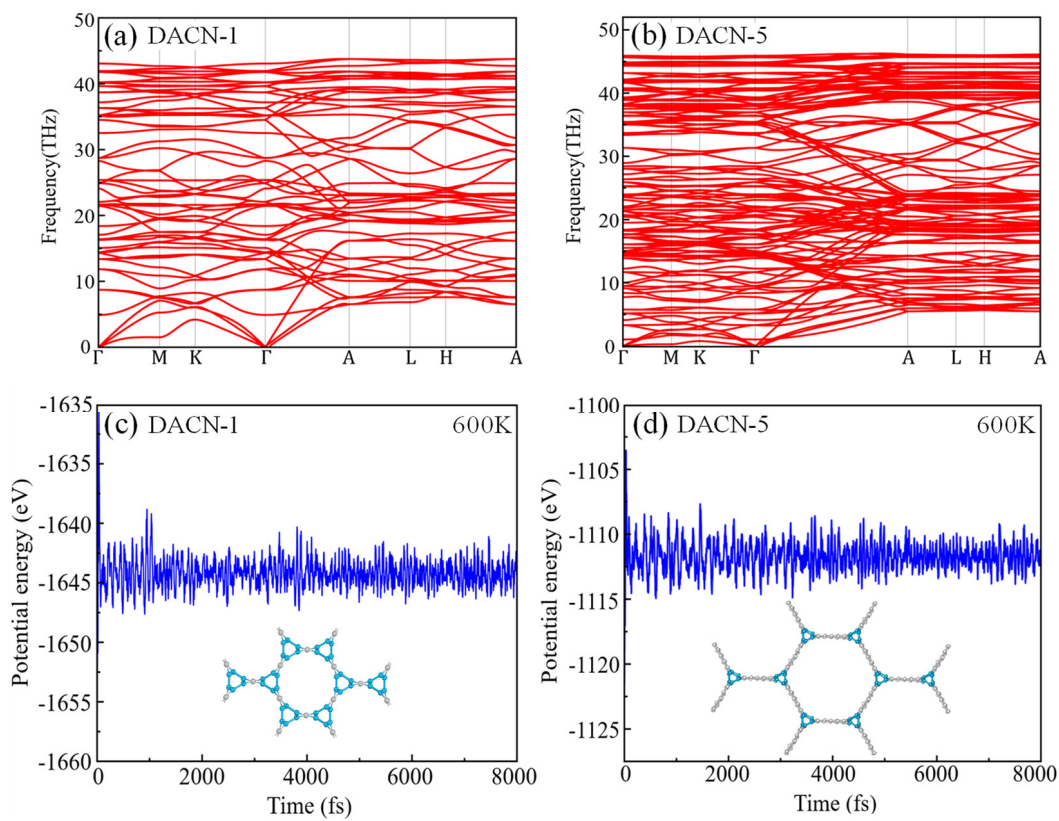

FIG. 2. Dynamical and thermal stabilities of DACN-1 and DACN-5. (a-b) Phonon dispersion for DACN-1 and DACN-5 in the entire BZ. (c-d) Total potential energy fluctuation for DACN-1 and DACN-5 during AIMD simulation at 600 K. The insets show the equilibrium structures of DACN-1 and DACN-5 after 8000 fs of AIMD simulation at 600 K.

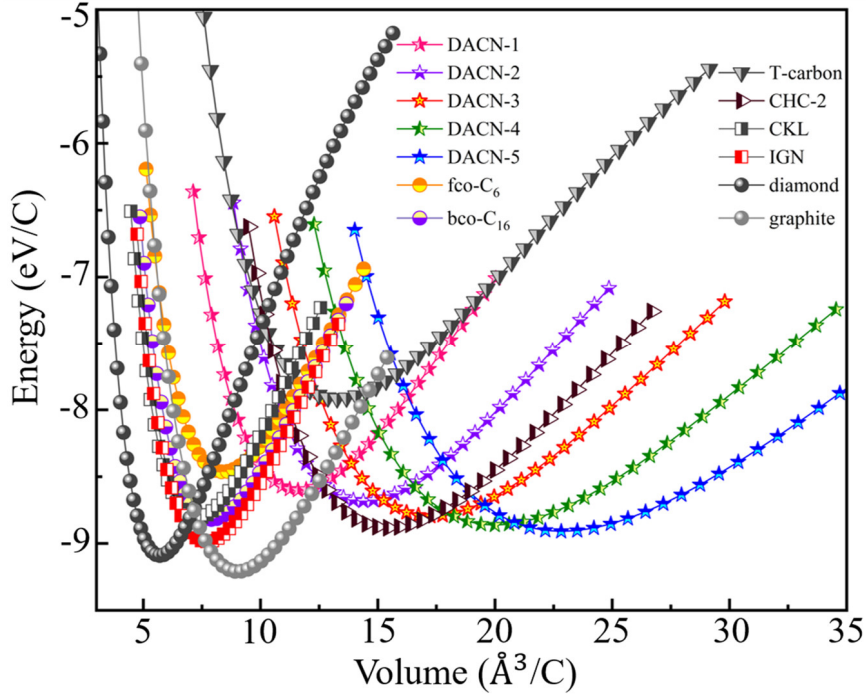

FIG. 3. Energy-volume plots for DACN-$n$ ($n$ = 1-5), diamond, graphite, T-carbon, CHC-2, fco-$C_6$, bco-$C_{16}$, CKL, and IGN.

The most interesting property of the DACN-$n$ family lies in its electronic properties. Due to the structural similarity of the triangulated DAGNRs (blue atoms) and FAGNRs (grey atoms) with varying widths ($n$), the electronic properties of DACN-$n$ are similar across different structures [see Fig. 1]. In the following discussion, we focus on the $n$ = 1 and $n$ = 5 structures, while the detailed electronic properties for the $n$ = 2 to $n$ = 4 structures are shown in Figs. S3-S5 of the SM. Figure 4(a-b) displays the band structures along the high-symmetry paths and the projected density of states (PDOS) of DACN-1. We can find that there exist only two non-degenerate energy bands that cross each other along the Γ-M and K-Γ paths, resulting in two double degeneracy points, labeled $W_1$ and $W_2$. These points are located at 2.7 meV and 3.8 meV above the Fermi level, respectively [see Table II], thus indicating the presence of an ideal Weyl semimetal. Additionally, the PDOS shows that the dominant contributions near the Fermi level come from the $p_x/p_y$ orbitals, similar to the $p_z$ orbitals in graphene. This can also be seen from the charge density distributions of the two states, $P_1$ and $P_2$, around the degenerate point [see Fig. 4(g)]. For DACN-$n$ ($n = 2,3,4$), it can be observed that their PDOS and charge density distributions are similar to those of DACN-1 [see Figs. S3-S5 in the SM].

Since the SOC effect is very weak in this system, the electron spin of DACN-$n$ can be treated as a dummy degree of freedom. Consequently, the crossing points ($W_1$ and $W_2$) formed by the intersection of two non-degenerate bands at the Fermi level, represent perfect Weyl points (WPs). It is noteworthy that the point group of DACN-$n$ family is $D_6$, and the absence of spatial inversion ($P$) and mirror symmetry results in neither $PT$ symmetry nor mirror symmetry protection for nodal lines. Therefore, when two energy bands cross, the intersection points manifest as discrete pairs of Weyl points. To confirm this, we conducted a thorough search for the intersection points of the two crossing bands across the entire Brillouin zone, identifying six pairs of Weyl points in the $k_z = 0$ plane. To further determine the topological charges associated with these Weyl points, we calculated the band dispersion of $W_1$ and $W_2$ along the $k_z$ direction [see Figs. 4(c-d)]. The results reveal that the crossing bands near the Fermi level exhibit linear dispersion along the $z$-axis, consistent with the characteristics of WPs associated with a Chern number of $C = \pm 1$. Additionally, we perform the integration of Berry curvature over a sphere enclosing the WPs [see Fig 4(e)], yielding a topological charge of -1 for $W_1$ along the Γ-M path and +1 for $W_2$ along the K-Γ path [inset of Fig 4(a)]. These results confirm the existence of two inequivalent Weyl points ($W_1$ and $W_2$) with a topological charge of $\pm 1$. Furthermore, the topological properties of the DACN-$n$ ($n = 2,3,4$) are similar to these observed in DACN-1 [see Figs. S3- S5 in the SM]. All six pairs of Weyl points in DACN-$n$ ($n = 2,3,4$) are distributed in the $k_z = 0$ plane within the BZ. Our integration of the Berry curvature shows that $W_1$ along the Γ-M path has a topological charge of -1, and $W_2$ along the K-Γ path has a topological charge of +1 [see Fig. S3-5(b)], which is consistent with the findings for DACN-1.

TABLE II. Energies, positions, topological charges and multiplicities of the two inequivalent Weyl points ($W_1$ and $W_2$) of DACN-$n$ ($n$ = 1-5).

| Material | WP | E (meV) | Coordinate ($k_1, k_2, k_3$) | Charge | Multiplicity |
|---|---|---|---|---|---|
| DACN-1 | $W_1$ | 2.7 | (0.275, 0, 0) | −1 | 6 |
|  | $W_2$ | 3.8 | (0.159, 0.159, 0) | +1 | 6 |
| DACN-2 | $W_1$ | 2.0 | (0.479, 0, 0) | −1 | 6 |
|  | $W_2$ | -0.1 | (0.250, 0.250, 0) | +1 | 6 |
| DACN-3 | $W_1$ | -17.8 | (0.031, 0.031, 0) | −1 | 6 |
|  | $W_2$ | -17.8 | (0.054, 0, 0) | +1 | 6 |
| DACN-4 | $W_1$ | 0.4 | (0.133, 0.133, 0) | −1 | 6 |
|  | $W_2$ | -0.6 | (0, 0.231, 0) | +1 | 6 |
| DACN-5 | $W_1$ | 1.3 | (0.421, 0.159, 0) | −1 | 6 |
|  | $W_2$ | 1.4 | (0.270, 0.270, 0) | +1 | 6 |

A key characteristic of WSMs is the presence of surface Fermi arcs that connect opposite chiral Weyl fermions. To investigate this feature, we have calculated the surface band structure of DACN-1 on the (001) surface along the projected $\bar{M} - \bar{\Gamma} - \bar{K}$ path [see Fig. 4(f)]. It is evident that $W_1$ and $W_2$ are projected onto the $\bar{\Gamma} - \bar{M}$ and $\bar{\Gamma} - \bar{K}$ paths on the (001) surface, respectively. The surface state manifests as a Fermi arc connecting a pair of opposite chiral WPs ($W_1$ and $W_2$). This extensive Fermi arc makes DACN-1 easy to be observed experimentally. Additionally, the surface states of DACN-$n$ ($n = 2,3,4$) also exhibit even longer Fermi arcs connecting pairs of opposite chiral Weyl points compared to those in DACN-1.

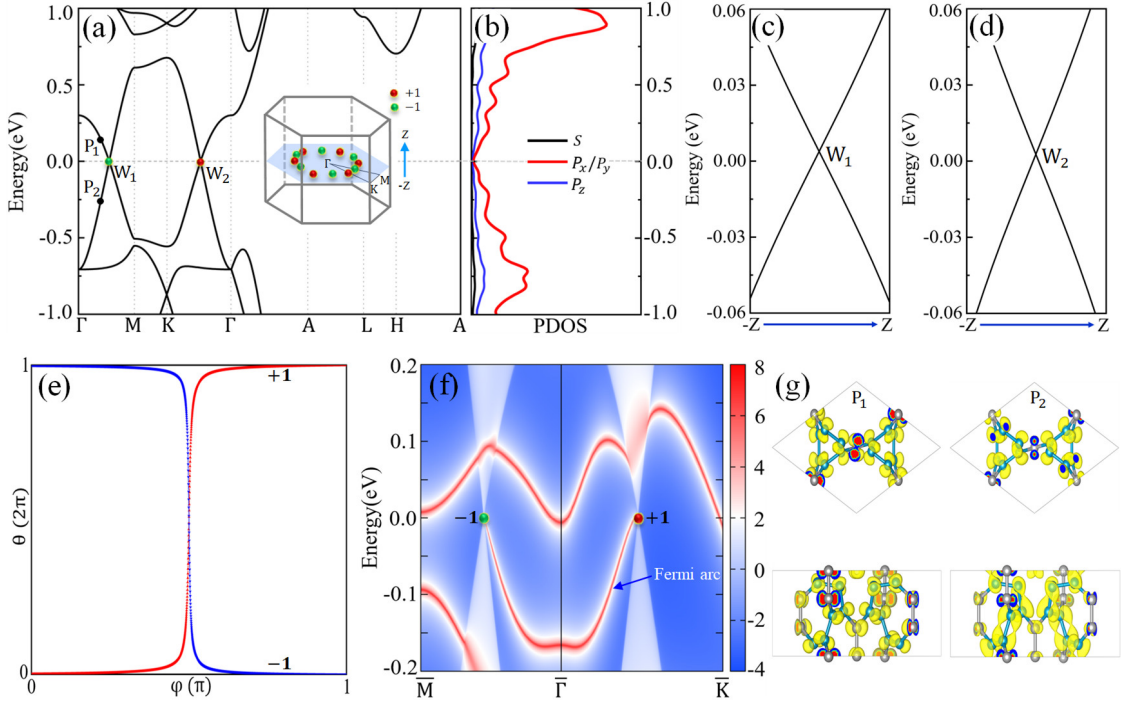

FIG. 4. Band structure, projected density of states (PDOS), chirality, surface band structure, and charge density of DACN-1. (a) Band structure along the high-symmetry paths and (b) PDOS of DACN-1. The 12 Weyl Points (WPs) distributed on the $k_z = 0$ plane in the BZ is show in the inset. (c) and (d) display the band dispersions along the $k_z$ direction for $W_1$ and $W_2$ of DACN-1, respectively. (e) Evolution of the Wannier charge centres of $W_1$ and $W_2$, where θ and φ denote the azimuthal angles in different directions around the closed sphere of the Weyl point. (f) Surface band structure on the semi-infinite (001) surface of DACN-1. (g) Top and side views of the charge densities of the $P_1$ and $P_2$ states around the WPs.

The band structure of DACN-5 is displayed on the Fig. 5(a). It is clear that two WPs ($W_1$ and $W_2$) exist along the M − K and K − Γ paths, respectively. The Weyl points are located above the Fermi energy level at 1.3 meV and 1.4 meV, respectively [see Table II], indicating the presence of perfect Weyl points. The chirality of $W_1$ along the M-K path is calculated to be C = -1, while that of $W_2$ along the K-Γ path is C = +1 [see Fig. 5(e)]. A comprehensive scan of the entire BZ reveals a total of 12 WPs in the $k_z = 0$ plane [see the inset of Fig. 5(a)]. Notably, the bulk energy band structure of DACN-5 differs significantly from that of DACN-1, displaying double Kagome bands. These include a red Kagome band ranging from [-0.65, 0.62] eV and a blue Kagome band from [-0.43, 0.62] eV, both of which consist of flat and Dirac-type bands. These

WPs originate from intersections between the Dirac-type bands in the two sets of Kagome bands. At the same time, we calculated the projected density of states (PDOS) [see Fig. 5(b)], revealing that the predominant contributions near the Fermi level come from the $p_x/p_y$ orbitals, similar to the $p_z$ orbitals in graphene. Figure 5(g) shows the charge densities for four states in the two sets of Kagome bands, The $P_1$ and $P_2$ states in the red Kagome band are primarily associated with $p_x/p_y$ orbitals, while the $P_3$ and $P_4$ states in the blue Kagome band include mainly contributions from $p_z$ orbitals. Additionally, one can see that the surface state of the surface band structure projected on the (001) surface [see Fig. 5(f)] shows Fermi arcs that connect a pair of opposite chiral WPs. This novel carbon material, which combines ideal Weyl semimetals with double Kagome bands, opens new avenues for investigating the interplay between ideal Weyl fermions and Kagome bands.

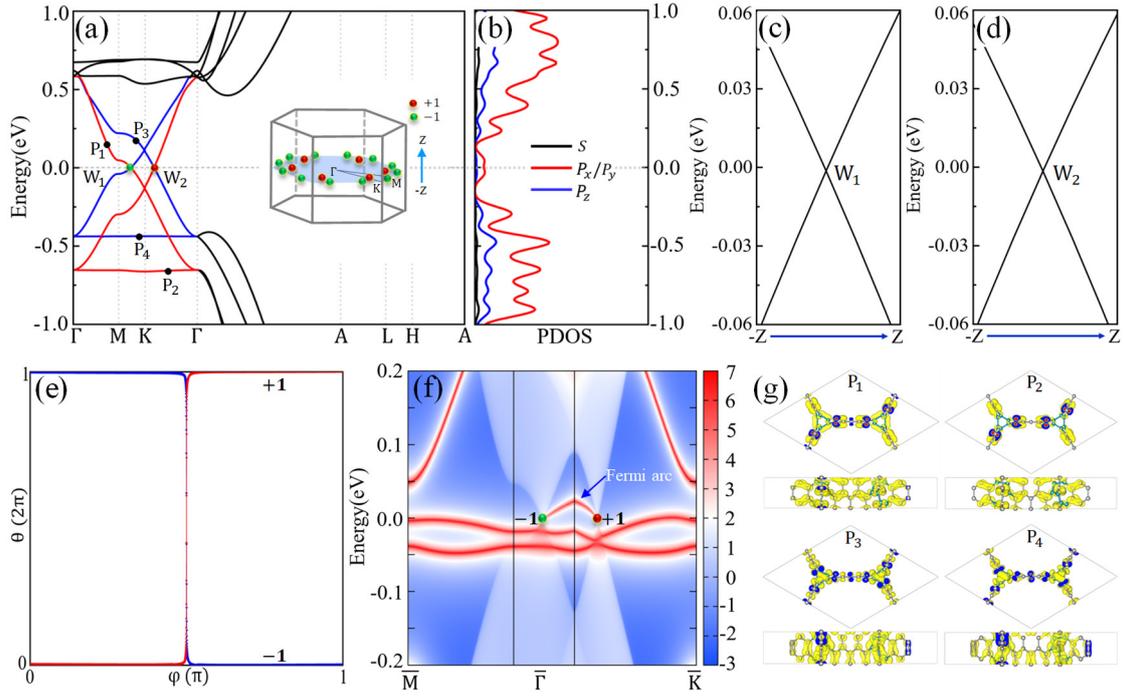

FIG. 5. Band structure, projected density of states (PDOS), chirality, surface band structure and charge density of DACN-5. (a) Band structure along the high-symmetry paths and (b) PDOS of DACN-5. The inset shows 12 WeyI points (WPs) distributed on the $k_z = 0$ plane in the BZ. The red and blue band structures exhibit double Kagome bands. (c) and (d) display the band dispersions along the $k_z$ direction for $W_1$ and $W_2$ of DACN-5, respectively. (e) Evolution of the Wannier charge canters of $W_1$ and $W_2$, where θ and φ denote the azimuthal angles in different directions around the closed sphere surrounding the Weyl point. (f) Surface band structures on the semi-infinite (001) plane of DACN-5. (g) Top and side views of the charge densities of the four different states ($P_1$, $P_2$, $P_3$, and $P_4$).

To guide experimental identification of DACN-$n$, we simulated the x-ray diffraction (XRD) spectra of DACN-$n$ for comparison with T-carbon, graphite and diamond [see Fig. S6 in the SM]. The (100) main peaks of DACN-$n$ exhibit a pronounced leftward shift with increasing width $n$, while the secondary peaks at (110), (200), and (210) gradually approach the main peak as $n$ increases. This distinctive behavior enhances the recognizability of DACN-$n$ in experimental observations. Importantly, the weak peak observed on the (210) facet of DACN-2 corresponds to a weak peak at 2θ slightly below 30° in the XRD spectra of chimney ash, indicating a potential presence of DACN-2 in chimney ash and providing a practical avenue for its experimental detection.

Here, we wish to highlight the significance and advantages of our current research. First, the DACN-$n$ family possesses a porous structure that offers a large specific surface area, enhancing its performance across various applications, including catalysis, adsorption, and energy storage. Its low density and small bulk modulus, derived from its porous nature, indicate promising potential in aerospace and cosmic materials. Second, compressing the basic structural units of DACN-$n$ into a single point leads to the formation of a Kagome lattice in a 2D plane. Notably, DACN-5 reveals a double Kagome band structures near the Fermi level, marking the first reported instance of such a structure in 3D bulk materials. If the flat bands within the Kagome band structure can be shifted to the Fermi level, this may lead to intriguing properties, such as superconductivity, ferromagnetism, and Wigner crystals [60, 61]. Finally, despite the discovery of numerous TSMs, ideal TSMs featuring nontrivial band intersections at the Fermi level are still relatively rare, especially in the context of ideal Weyl semimetals. Thus, the DACN-$n$ family emerges as a highly anticipated series of ideal Weyl semimetals, providing new insights into the search for such materials among lightweight materials with negligible SOC effect.

## 4. Conclusions

In conclusion, we employed first-principles calculations and symmetry analysis to predict a series of fully $sp^2$-hybridized carbon networks composed of distorted armchair-type graphene nanoribbons, termed DACN-*n*. The structural stability of DACN-*n* is confirmed by phonon spectrum, *ab initio* molecular dynamics simulations, and elastic constant calculations. Notably, we find that it only contains two non-degenerate bands that intersect to form twofold degenerate points near the Fermi level, resulting in perfect six pairs of Weyl points with Chen number $C = \pm 1$ distributed on the $k_z = 0$ plane. Additionally, DACN-5 exhibits a unique double Kagome band near the Fermi level, and its twofold degenerate Weyl points are derived from the intersection between Dirac-type bands in the two sets of Kagome bands. We further explore the surface band structure on the (001) surface, where surface states manifest as segments of Fermi arcs connecting pairs of opposite chiral Weyl points. Finally, to guide experimental identification, we simulated x-ray diffraction (XRD) experiments. Our work not only contributes a new class of members to the family of 3D carbon allotropes, but also offers an excellent platform for investigating ideal Weyl fermions and double Kagome bands.

## CRediT authorship contribution statement

Y.Y.B. carried out the DFT calculations, and writing-original draft preparation. Y.G. conceived and designed the project. W.W. and Z.-Y.L. analyzed the results. Y.G and Y.L. writing-reviewing, conceptualization, supervision, and project administration.

## Declaration of competing interest

The authors declare that they have no known competing financial interests or personal relationships that could have appeared to influence the work reported in this paper.

## Acknowledgements

We wish to thank Quansheng Wu and Chengyong Zhong for helpful discussions. This work was supported by the National Natural Science Foundation of China (Grants No. 12304202), Hebei Natural Science Foundation (Grant No. A2023203007), Science Research Project of Hebei Education Department (Grant No. BJK2024085), and Shandong Provincial Natural Science Foundation (Grant No. ZR2023QA012).